# Self-Contained Cross-Cutting Pipeline Software Architecture

## Amol Patwardhan[1], Rahul Patwardhan[2], Sumalini Vartak[3]

[1]Software Architect I, CRIF Corporation, LA, USA
[2]IT Technical Manager, Infobahn Softworld Inc, CA, USA
[3]Senior Software Engineer, Lead IT, IL, USA

**Abstract -** *Layered software architecture contains several intra-layer and inter-layer dependencies. Each layer depends on shared components making it difficult to release a code change, bug fix or feature without exhaustive testing and having to build the entire software code base. This paper proposed self-contained, cross-cutting pipeline architecture (SCPA) that is independent of existing layers. We chose 2 open source projects and 3 internal intern projects that used n-tier architecture and applied the SCPA to release subsequent feature additions and any bug fixes. The SCPA decreased the release time by 42.99%. The lines of delivered code (LOC), increased by 22.58%. The number of defects found in existing functionality decreased by 85.54%. The SCPA also provided ability to roll back or switch off the new feature quickly. SCPA proved a suitable architecture for agile software development and continuous deployment.*

## 1. INTRODUCTION

Layered software architecture is a highly popular choice of implementation in the software industry. The user interface layer consists of fields, controls with which users interact. The middle tier processing layer consists of business logic code and business specific entities. The data layer consists of code that interacts with the database. Such a layered architecture comes in different flavours such as n-tier architecture, 3-tier architecture and model view controller (MVC) architecture.

One of the design principles and best practices in software development is software reuse. This leads to creation of shared components (dlls, api, libraries, resource files, services) that span across various layers of the software architecture. For instance, the user interface layer consists of several utilities to facilitate data type conversion, data formatting and static utility classes for rendering controls. Similarly, the business layer consists of complex shared libraries to manage calling into the data layer, execute file input/output operation, and perform report rendering or processing asynchronous backend jobs. The data layer also contains dedicated classes (data access layer controllers) to handle the creation of records, reading data, update and deletion of data which is commonly called CRUD operation. As part of the layered software architecture a component from higher layer is allowed to access a component in a lower layer only if the component has the same responsibility and context. For example, the business controller code for Product is allowed to use the data controller designated to execute CRUD on product data and any related tables. This means, if the product business controller needs to access sales data it has to include and call into the sales business controller which in turn calls into the corresponding sales data controller.

As software matures the interdependency on components residing in same layer, to access the necessary data, increases. This leads to a monolithic tightly coupled intra-layer interdependent matrix of several components with single responsibilities. A problem arising from this dependency is that any subsequent changes, bug fix or new feature additions cannot be developed and released independently. The entire software has to be built and regression tested to ensure no performance side effects occur, no bugs are introduced and no changes in functionality are introduced. The dependencies result in a lengthy delivery time, higher testing effort and increase in development cost.

This paper proposes self-contained cross-cutting pipeline architecture (SCPA) to minimise the interdependency across various layers of software and improve the efficiency of software maintenance and release effort. The new architecture SCPA would allow faster development, testing and deployment without having to compile the entire software code base. The SCPA would reduce the testing effort and enable deploying or rolling back the code change, bug fix without affecting other parts of the system.

## 1.1 Related Work

The feasibility of N-Tier layered architecture in an increasingly distributed environment in terms of deployment complexity was evaluated by Manuel [1]. The study recommended a data centric design instead of moving code or data for message based interaction. A detailed study [2] on various software architectures has been done. The review discussed advances in architecture optimization for maintaining quality requirements. N-Tier layered architecture has been used for implementation of windows desktop applications and interfacing with embedded devices and sensors [3], [4], [5]. Shaw [6] showed that improvements in technology and tools have resulted in efficient software development processes. Kruchten [7] discussed the past, present and future evolution of software architecture practices. Nord [8] evaluated the strengths and weaknesses of agile and traditional software architecture process. The research recommended distinguishing between plan-based software developments process from the architecture.

A comparative study [9] was done between various software methodologies, architecture designs such as Rapid Application Development (RAD), service oriented architecture. A research [10] done on optimizing software architecture design proposed techniques like software decomposition, tool framework for deployment and other improvements for better quality and efficient release process.

## 1.2 SCPA

We proposed self-contained, cross-cutting pipeline architecture (SCPA) to solve the problem of intra and inter layer dependency in a software code base. The architecture is self-contained because the implementation contains code for all the necessary business requirements and does not depend on any shared code from other layers. The self-contained code can be easily built and deployed without having to rebuild the whole software code base. The architecture is cross-cutting because the implementation contains code that belongs to user interface layer, business layer and even data access layer if needed and thus cuts across several layers of the architecture.

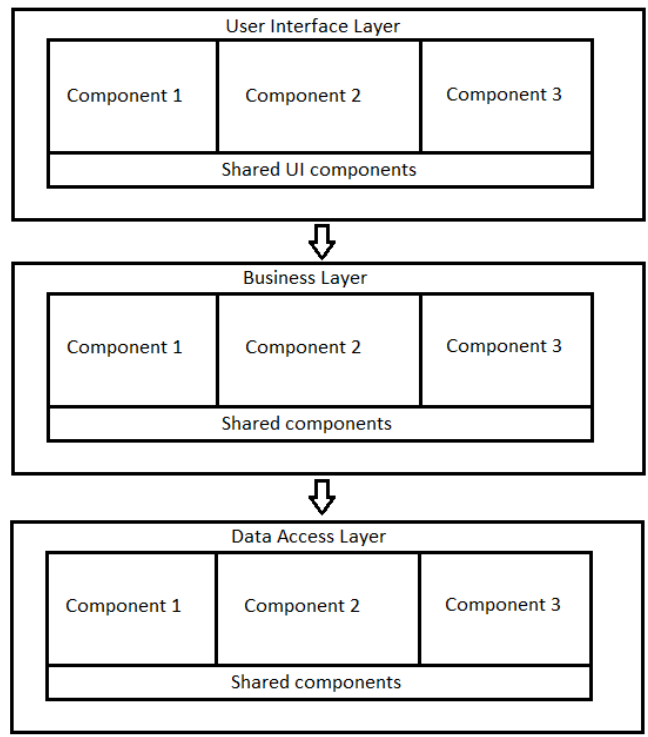

**Fig -1**: Traditional N-Tiered Architecture

The figure above shows the typical n-tiered layered architecture and the interdependencies. Assume that component 1 handles product data and component 3 handles sales data. To show the sales by product in the UI it is evident that component 1 needs to rely on component 3 for CRUD operations related to sales data. If there is a bug in the processed product sale data or a new feature needs to be added then all the interdependent components have to be compiled, tested and deployed.

On the contrary the proposed SCPA architecture completely removes these dependencies by decoupling the components and including them in a separate pipeline of their own. This self-contained, cross-cutting pipeline structure is shown in Figure 2. The U1, B1 and D1 are sub-components of the same pipeline that cuts across several layers without depending on any other component. U1, U2 represent User Interface Layer specific code. B1, B2 represents business layer specific code. D1, D2 represents data access layer specific code.

Each pipeline can be loaded in separate runtime context using a plug-in interface. The legacy code needs to declare an interface called i-plugin which creates 3 methods: a) Load b) Execute c) Next. This interface serves as a contract for concrete implementations residing in each pipeline.

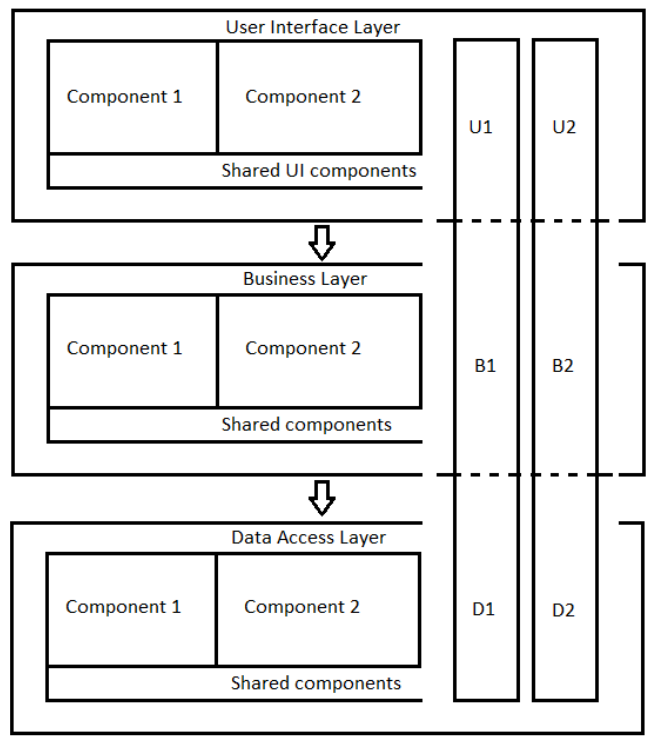

**Fig -2**: Self-Contained Cross-Cutting Pipeline Architecture (SCPA)

Each sub-component in the pipeline should implement these methods. This structure allows the legacy code to load and execute the pipeline code in a processing chain. It also decouples the legacy code from any new code added to the code base or any bug fix. Only the new assembly (dll) has to build and deployed making the code deployment straightforward. To switch the fix or feature on/off, the dll can be simply deployed or removed from a folder.

## 2. METHODOLOGY

We first used a total of 5 projects (2 open source and 3 internal) to obtain the baseline measurements for our study. The first project was a library management system (Project 1) and the second project was a user forum (Project 2). These projects had been implemented using Microsoft technology stack consisting of c#, ASP.NET and MS SQL as the database. The first project was a windows desktop application. The second project was a web application. In addition to the open source project code base we also used 3 internal projects completed by student interns. These three projects were a task manager program (Project 3), a bug tracking program (Project 4) and a file parser program (Project 5). As a result, we were able to obtain our baseline readings from a wide range of design and architecture styles. We were able to get baseline from a windows application, web application, project using web forms, MVC, project using database and project without a database. The readings were obtained over a period of 15 months and contained multiple feature releases and maintenance cycles.

**Table -1:** Baseline readings from 5 candidate projects

|  | P1 | P2 | P3 | P4 | P5 |
|---|---|---|---|---|---|
| Avg. Post Release Defect | 1.4 | 1.13 | 0.87 | 1 | 1.13 |
| Average Release Time | 28.6 | 27.73 | 7 | 8.73 | 6.87 |
| Average Testing Time | 7.4 | 7.6 | 1.33 | 2.4 | 2.07 |
| Average Development Time | 7.13 | 7 | 2.53 | 2.4 | 2.53 |
| Average Deployment Time | 3.07 | 4.46 | 0.93 | 1.2 | 1.13 |
| Average LOC Changed | 1715.33 | 407.67 | 13,27 | 19.6 | 16.27 |

Once the baseline results were obtained any subsequent feature development, code change, bug fix was implemented using the SCPA. The development, testing, deployment, post release bug count and lines of code delivered were measured over a period of another 15 months.

**Table -2:** SCPA results from 5 candidate projects

|  | P1 | P2 | P3 | P4 | P5 |
|---|---|---|---|---|---|
| Avg. Post Release Defect | 0.2 | 0.2 | 0.13 | 0.13 | 0.13 |
| Average Release Time | 20.2 | 16.93 | 2.6 | 2.4 | 2.87 |
| Average Testing Time | 4.93 | 4.87 | 0.6 | 1.3 | 0.8 |
| Average Development Time | 2.67 | 3.87 | 1.27 | 1.3 | 1 |
| Average Deployment Time | 1.07 | 1.33 | 0.93 | 0.53 | 0.46 |
| Average LOC Changed | 2019.9 | 568.2 | 17.8 | 33.33 | 23.4 |

The results indicated that there was an overall decrease in the average post release defect, average release time, average development time, average testing time and an increase in average lines of code delivered per release.

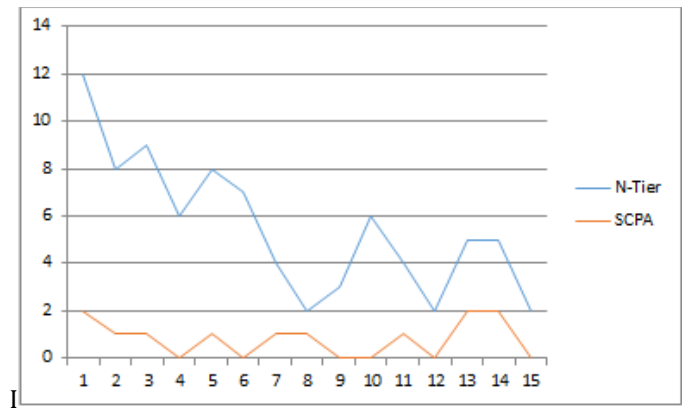

**Chart -1**: Post Release Defect Count

The above chart shows that the number of defects found using SCPA was lesser than traditional N-Tier architecture implementations.

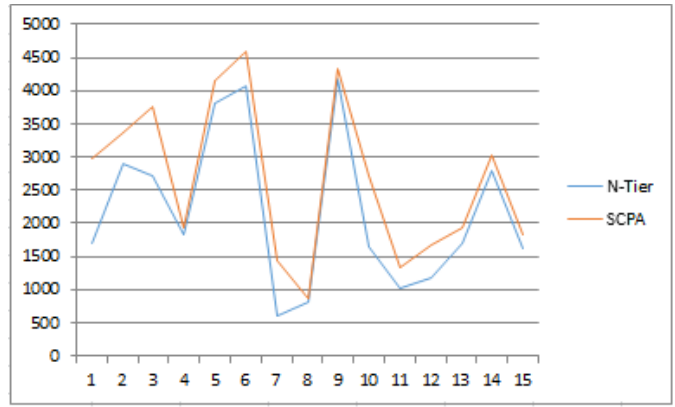

**Chart -2**: Lines of Code

The above chart shows that the number of lines of code delivered per release was more in case of SCPA.

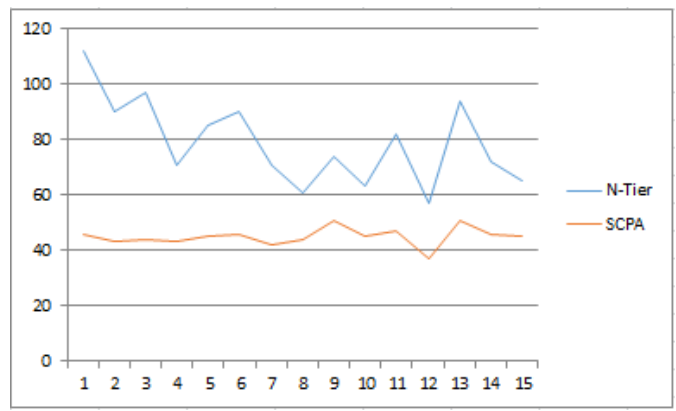

**Chart -3**: Release Duration in Days

The above chart shows that the release duration in days was lesser than traditional N-Tier architecture implementations, when SCPA was applied. It can be seen that the overall release duration decreased by 42.99%. This showed that the SCPA improved the software delivery process efficiency. Moreover, the post release defect count decreased by 85.54% which indicates that the software quality improved when SCPA was used and thus was an indicator of lower software maintenance cost. Finally, the average lines of code delivered increased by 22.58% which means the development efficiency increased when SCPA was used and more features, bugs and code changes were released per available development resource.

## 3. CONCLUSIONS

In this paper we proposed a new self-contained and cross –cutting pipeline architecture to solve the problem of interdependencies across architectural layers (intra and inter layer) and dependencies on shared libraries, api, and utility code. The proposed architecture eliminated the need to compile the entire code base. Any code changes were completely self-contained and could be quickly released without affecting other parts of the software product. The features and code changes made using SCPA could be easily switched on/off or deployed or rolled back. The SCPA architecture drastically reduced development, testing and deployment time and effort and is highly suitable for continuous deployment and agile software development.